\documentclass[twocolumn]{aastex61}
\usepackage{amsmath,amstext}
\usepackage[T1]{fontenc}
\usepackage{apjfonts} 
\usepackage[figure,figure*]{hypcap}

\newcommand{\name}{1I/`Oumuamua} 

\received{2017 November 16}
\revised{2017 December 4}
\accepted{2017 December 6}
\submitjournal{ApJL}

\shorttitle{Col-OSSOS: Colors of 1I/`Oumuamua}
\shortauthors{Bannister et al.}

\begin{document}

\title{Col-OSSOS: Colors of the Interstellar Planetesimal 1I/`Oumuamua}

\author[0000-0003-3257-4490]{Michele T. Bannister}
\correspondingauthor{Michele T. Bannister}
\email{michele.t.bannister@gmail.com}
\affiliation{Astrophysics Research Centre, School of Mathematics and Physics, Queen's University Belfast, Belfast BT7 1NN, United Kingdom}

\author[0000-0003-4365-1455]{Megan E. Schwamb}
\affiliation{Gemini Observatory, Northern Operations Center, 670 North A'ohoku Place, Hilo, HI 96720, USA}

\author[0000-0001-6680-6558]{Wesley C. Fraser} 
\affiliation{Astrophysics Research Centre, School of Mathematics and Physics, Queen's University Belfast, Belfast BT7 1NN, United Kingdom}

\author[0000-0001-8617-2425]{Michael Marsset}
\affiliation{Astrophysics Research Centre, School of Mathematics and Physics, Queen's University Belfast, Belfast BT7 1NN, United Kingdom}

\author[0000-0003-0250-9911]{Alan Fitzsimmons} 
\affiliation{Astrophysics Research Centre, School of Mathematics and Physics, Queen's University Belfast, Belfast BT7 1NN, United Kingdom}

\author{Susan D. Benecchi}
\affiliation{Planetary Science Institute, 1700 East Fort Lowell, Suite 106, Tucson, AZ 85719, USA}

\author[0000-0002-1708-4656]{Pedro Lacerda} 
\affiliation{Astrophysics Research Centre, School of Mathematics and Physics, Queen's University Belfast, Belfast BT7 1NN, United Kingdom}

\author[0000-0003-4797-5262]{Rosemary E. Pike}
\affiliation{Institute for Astronomy and Astrophysics, Academia Sinica; 11F AS/NTU, National Taiwan University, 1 Roosevelt Rd., Sec. 4, Taipei 10617, Taiwan}

\author[0000-0001-7032-5255]{J.~J. Kavelaars}
\affiliation{Herzberg Astronomy and Astrophysics Research Centre, National Research Council of Canada, 5071 West Saanich Rd, Victoria, British Columbia V9E 2E7, Canada}
\affiliation{Department of Physics and Astronomy, University of Victoria, Elliott Building, 3800 Finnerty Rd, Victoria, BC V8P 5C2, Canada}

\author{Adam B. Smith}
\affiliation{Gemini Observatory, Northern Operations Center, 670 North A'ohoku Place, Hilo, HI 96720, USA}

\author{Sunny O. Stewart}
\affiliation{Gemini Observatory, Northern Operations Center, 670 North A'ohoku Place, Hilo, HI 96720, USA}

\author{Shiang-Yu Wang}
\affiliation{Institute of Astronomy and Astrophysics, Academia Sinica; 11F of AS/NTU Astronomy-Mathematics Building, Nr. 1 Roosevelt Rd., Sec. 4, Taipei 10617, Taiwan}

\author[0000-0003-4077-0985]{Matthew J. Lehner}
\affiliation{Institute of Astronomy and Astrophysics, Academia Sinica; 11F of AS/NTU Astronomy-Mathematics Building, Nr. 1 Roosevelt Rd., Sec. 4, Taipei 10617, Taiwan}
\affiliation{Department of Physics and Astronomy, University of Pennsylvania, 209 S. 33rd St., Philadelphia, PA 19104, USA}
\affiliation{Harvard-Smithsonian Center for Astrophysics, 60 Garden St., Cambridge, MA 02138, USA}

\begin{abstract}

The recent discovery by Pan-STARRS1 of 1I/2017 U1 (`Oumuamua), on an unbound and hyperbolic orbit, offers a rare opportunity to explore the planetary formation processes of other stars, and the effect of the interstellar environment on a planetesimal surface.
1I/`Oumuamua's close encounter with the inner Solar System in 2017 October was a unique chance to make observations matching those used to characterize the small-body populations of our own Solar System. 
We present near-simultaneous g$^\prime$, r$^\prime$, and J photometry and colors of 1I/`Oumuamua from the 8.1-m Frederick C. Gillett Gemini North Telescope,
and $gri$ photometry from the 4.2 m William Herschel Telescope.
Our g$^\prime$r$^\prime$J observations are directly comparable to those from the high-precision \textit{Colours of the Outer Solar System Origins Survey} (Col-OSSOS), which offer unique diagnostic information for distinguishing between outer Solar System surfaces. 
The J-band data also provide the highest signal-to-noise measurements made of 1I/`Oumuamua in the near-infrared.
Substantial, correlated near-infrared and optical variability is present, with the same trend in both near-infrared and optical. 
Our observations are consistent with 1I/`Oumuamua rotating with a double-peaked period of $8.10 \pm 0.42$ hours and being a highly elongated body with an axial ratio of at least 5.3:1, implying that it has significant internal cohesion. 
The color of the first interstellar planetesimal is at the neutral end of the range of Solar System $g-r$ and $r-J$ solar-reflectance colors: it is like that of some dynamically excited objects in the Kuiper belt and the less-red Jupiter Trojans.

\end{abstract}

\keywords{minor planets, asteroids: individual (1I/2017 U1)}

\section{Introduction} 
\label{sec:intro}

The first detection of an interstellar minor planet, 1I/2017 U1 (`Oumuamua), came on 2017 October 19, at $m_V = 19.6$, by Pan-STARRS1 \citep{Chambers.2016}) \citep{MPECU181,Meech:2017}.
In 19 years of digital-camera all-sky surveying, it is the first definitively interstellar, decameter-scale object to be found.
The earlier lack of detections has implied a low density of interstellar planetesimals \citep{Francis:2005br,Cook:2016,Engelhardt:2017}.
Several Earth masses of ejected bodies are expected per star during planetary formation and migration \citep[e.g.][]{Levison:2010,Barclay:2017}.
Given the number of Galactic orbits since the major ejection of planetesimals from the Solar System, \name\ is statistically unlikely to originate from the Solar System.

\name's orbit\footnote{\label{fn:JPL}JPL Horizons heliocentric elements, as of 2017 November 14: \url{https://ssd.jpl.nasa.gov/sbdb.cgi?sstr=2017\%20U1}} has a securely extrasolar origin. 
\name\ came inbound to the Sun on a hyperbolic and highly inclined trajectory that radiated from the solar apex, with an orbital eccentricity of $e=1.1994 \pm 0.0002$ and inclination of 122.7\degr, avoiding planetary encounters.  
Such orbits are not bound to our Solar System. 
The planetesimal was travelling at a startlingly high velocity of $v_{\infty} = 26.02 \pm 0.40$~km/s. 
This velocity is typical for the mean Galactic velocity of stars in the solar neighborhood \citep{Mamajek:2017arXiv}.

The physical properties of \name\ are not yet well constrained.
Its approach geometry and fast passage left only a brief window when it was observable, after its 0.16~au minimum approach to Earth on 2017 October 14.
As the NEOWISE scans missed its outbound trajectory (J. Masiero, pers. comm.), no albedo is yet known. 
\name\ has an absolute magnitude of $H_V = 22.08 \pm 0.45$\added{\footnote{See footnote 1.}}, implying a size of $\lesssim 200$~m, assuming it has an albedo in the range seen for either carbonaceous asteroids or Centaur albedos of $p_V = 0.06-0.08$ \citep{Bauer:2013,Nugent:2016}. 
No detection was seen 
in STEREO HI-1A observations (limiting magnitude of $m\sim13.5$) near \name's perihelion passage at 0.25~au on 2017 September 9 (K. Battams, pers. comm).
Consistent with this earlier non-detection, it was a point source in deep VLT imaging on 2017 October 24, with no coma \citep{MPECU183}, and upper limits of surface brightness of 28--30 mag arcsec$^{-2}$ within $\sim5\arcsec$ radial distance were set by \citet{Ye:2017arXiv} on October 26 and \citet{Knight:2017arXiv} on October 30.
Additionally, no meteor activity from associated dust was seen \citep{Ye:2017arXiv}.
This implies observations of \name\ directly measure its surface.

Measurement of the surface reflectivity of \name\ will provide the first ever comparison between solar planetesimals and those from another star.  
Such measurements could be used to infer the formation environment of this object, or provide evidence for a surface composition that is distinct from solar planetesimals. 
However, \name's surface composition may have experienced substantial alteration during its exposure of Myr, and potentially Gyr, in interstellar space.
No star has yet been confirmed as a potential origin \citep{Mamajek:2017arXiv}, therefore the upper bound on \name's age is around 10 Gyr, after the formation of stars of moderate metallicity.

Compositional information on minor planets as small as \name\ is limited.
In reflectance relative to the colour of the Sun, larger Solar System objects range from neutral to substantially more red \citep[e.g][and references therein]{Jewitt:2015}, with spectra ranging from featureless (C-type asteroids) to strong absorption bands (S-type asteroids) \citep{Rivkin:2015,Reddy:2015}.  
Initial observations of \name\ with 4--5-m class telescopes show a featureless spectral slope that is similar to many small trans-Neptunian objects (TNOs): moderately redder than solar.
Optical spectra in the wavelength range 400-950 nm from 2017 October 25 and 26 include slopes of $30 \pm 15\%$ \citep{Masiero:2017arXiv}, 
$17 \pm 2\%$ \citep{Fitzsimmons:2017}, 
and $10 \pm 6\%$ \citep{Ye:2017arXiv} per 1000 angstroms.

Both optical and near-infrared spectral information beyond $\sim 1 \mbox{ $\mu$m}$ are necessary to distinguish the compositional classes seen in the outer Solar System \citep{Fraser:2012cs,DalleOre:2015,Pike:2017gf}. 
Thus, J-band photometry is key for establishing the relationship of \name's surface type to the Solar System.
We present near-simultaneous $grJ$ photometry and colors of \name\ in the  optical and near-infrared, and compare to the colors of known Solar System bodies.

\section{Observations and Analysis}

\subsection{Observations}
\label{sec:observations}

\begin{figure}
\gridline{\fig{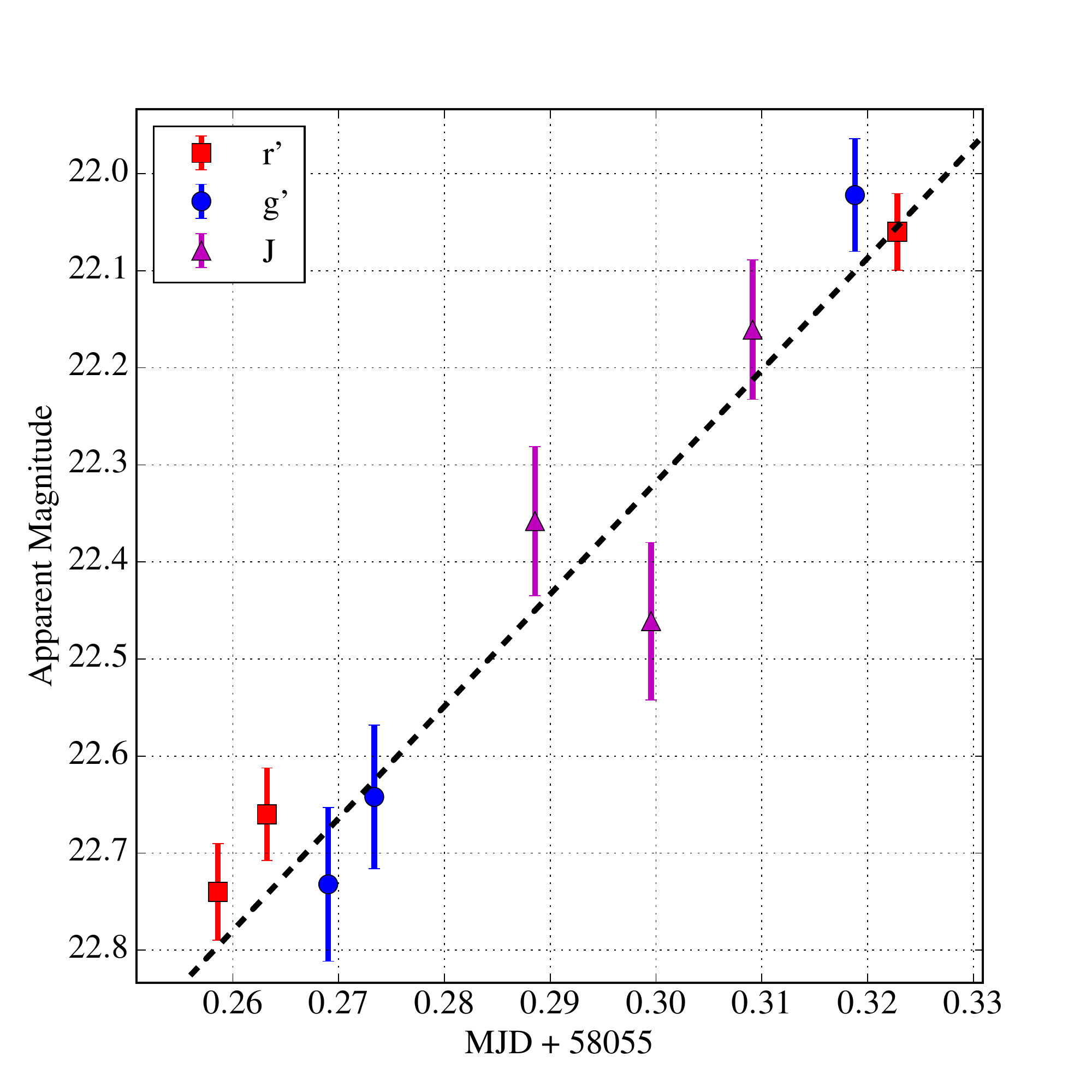}{\linewidth}{Color-corrected Gemini photometry of \name\ in g', r', and J (Table~\ref{tab:observations}). The line indicates best fit.
\label{fig:movingJband} }
}
\gridline{\fig{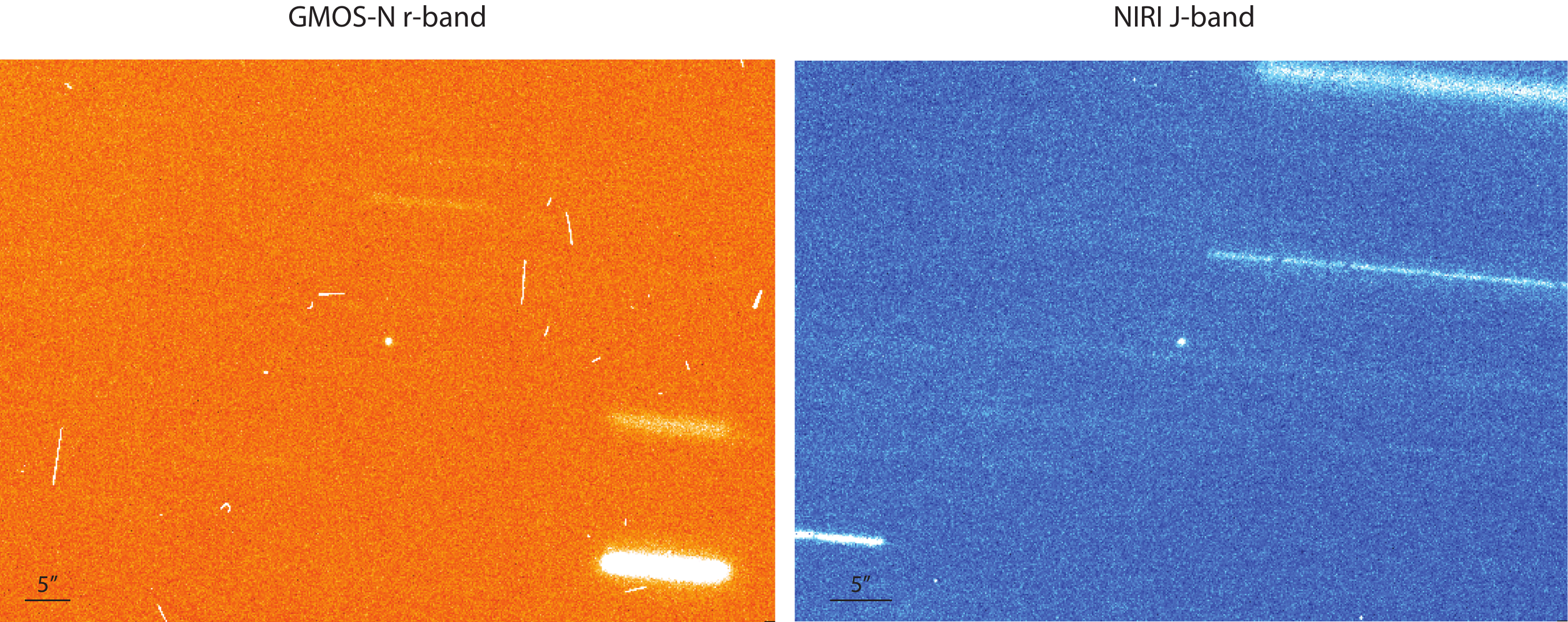}{\linewidth}{Imaging of \name\ with non-sidereal tracking on the R.A. 160\arcsec/hr motion rates of \name; a single 300 s GMOS exposure in $r$ (left) and a stack of 11 NIRI 120-s exposures in $J$ (right). \name\ was free of cosmic rays in all exposures.
\label{fig:imaging}}
}
\caption{Observations and photometry of \name\ from Gemini North on 2017 October 29.}
\end{figure}

We observed \name\ with two telescopes on 2017 October 29.
First, we observed \name\ with the 8.1-m Frederick C. Gillett Gemini North Telescope on Maunakea, during 05:50--07:55 UT.
JPL Horizons\footnote{\url{https://ssd.jpl.nasa.gov/horizons.cgi}} predicted from the available 15-day arc that \name\ was then at a heliocentric distance of 1.46 au and geocentric distance of 0.53 au (phase angle $\alpha = 24.0\degr$), producing a rapid rate of on-sky motion of R.A. 160 \arcsec/hr and Decl. 14 \arcsec/hr.
We therefore tracked the telescope non-sidereally at \name's rates (Fig.~\ref{fig:imaging}). 
The observations were in photometric skies between airmass 1.04-1.14, with seeing in r$^\prime$ of 0.7\arcsec\ to 0.5\arcsec.
The waxing 63\%-illuminated Moon was 39\degr\ away, producing a sky brightness of $\sim 20$ mag/arcsec$^{2}$ in r$^\prime$. 

g$^\prime$, r$^\prime$, and J imaging were obtained using the imaging mode of the Gemini Multi-Object Spectrograph \citep[GMOS-N,][]{2004PASP..116..425H} and the Near-Infrared Imager \citep[NIRI;][]{2003PASP..115.1388H}. 
The predicted apparent magnitude of \name\ was $m_v = 22.7$; exposure times are in Table~\ref{tab:observations}. 
We observed with GMOS in two filters: r$\_$G0303 ($\lambda$=6300 \AA, $\delta\lambda$=1360 \AA) and g$\_$G0301($\lambda$=4750 \AA, $\delta\lambda$=1540 \AA), which we refer to as r$^\prime$ and g$^\prime$, and which are similar to $r$ and $g$ in the Sloan Digital Sky Survey (SDSS) photometric system \citep{1996AJ....111.1748F}. 
GMOS was configured with the upgraded red-sensitive CCDs from Hamamatsu Photonics. 
The target was kept on the middle GMOS CCD, and 2$\times$2 binning was used, resulting in an effective pixel scale of 0\arcsec.1614.
NIRI J band ($\lambda$=12500 \AA, 11500-13300 \AA~coverage) images were acquired using the f/6 camera (pixel scale of 0\arcsec.116). 
For both instruments, we dithered between exposures in the same filter.

Any significant magnitude changes due to rotational variability of \name\ will affect its measured colors. 
Our observing program therefore employed the design of the Colours of the Outer Solar System Origins Survey \citep[Col-OSSOS; full detail of the techniques used will appear in][]{Schwamb:2017inprep}. 
Col-OSSOS uses repeated measurements in each band to distinguish any light curve effects from the change in surface reflectance during the observing sequence, bracketing the NIRI observations with GMOS imaging in a r$^\prime$g$^\prime$Jg$^\prime$r$^\prime$ filter pattern. 
With this cadence, we can identify if \name\ is variable on the timescale of our observations, and apply a correction, assuming that all filters are similarly affected (see \S~\ref{sec:colours}). 
This is reasonable as brightness variations for solar system objects in \name's size regime are due to object shape; use of filter bracketing during Col-OSSOS has been effective at removing light curve effects \citep{Pike:2017gf}.

We acquired specific observations for calibration temporally adjacent to our science data.
The optical sequence was bracketed by a single 150 s sidereally-tracked exposure in each optical filter, centered close to \name's predicted location.
For the NIR calibration, we observed NIR standard GD 246 at two different airmasses, with a set of nine dithered, sidereally tracked exposures at the beginning and end of our observing sequence. 

Thirteen hours after the Gemini observations, 
we observed \name, then at very similar geometry with a phase angle of $\alpha = 24.4\degr$, with the 4.2 m William Herschel Telescope (WHT) on La Palma from 19:45 UT -- 21:52 UT.
Non-sidereally guided imaging was obtained in photometric conditions and $\sim 1\arcsec$ seeing, at airmass 1.3--1.1, with the imaging mode of the ACAM imager/spectrograph \citep{Benn:2008} (pixel scale of 0\arcsec.253).
The data were acquired with four filters: ING Filter \#701, \#702, \#703 and \#704, corresponding 
to $g,r,i,z$ in the SDSS photometric system. Individual exposures were 100 s, in a sequence $6r-6g-7r-10i-6r-12z-6r$ (Table~\ref{tab:observations}), with the repetition of $r$ to test for variability.

\subsection{Data Reduction}
\label{sec:reduction}

The Gemini and WHT images\footnote{All reduced data are available at \url{http://apps.canfar.net/storage/list/ColOSSOS/Interstellar}.} were prepared for analysis using standard reduction techniques.
The GMOS observations were bias-subtracted and flat-fielded using standard methods with Gemini IRAF v1.14 in AstroConda, with a master flat frame built from the last month of GMOS twilight flats, and the CCD chips mosaiced into a single extension.
For the NIRI sequence, flat-fielding was with a master flat frame built from the Gemini facility calibration unit flats. 
A sky frame was produced for each image of \name\ from the unshifted 14 frames closest in time to the image, and subtracted from that image.
We removed cosmic rays from each NIRI image with L. A. Cosmic \citep{vanDokkum:2001}, then aligned and stacked 20 of the 21 science 120-s NIRI exposures, rejecting one frame where \name\ was in front of a background source. 
To assess for any variability in \name, we built independent stacks from each third of the J image sequence (Table~\ref{tab:observations}).
The WHT imaging were bias-subtracted and flatfielded with archival sky flats from 2017 August, as the sky flats taken prior to the observations had strong gradients across the field due to scattered moonlight and twilight variations.

Photometric measurements on all Gemini data (Table~\ref{tab:observations}) were performed with TRIPPy \citep{Fraser:2016a}, using a round aperture with a radius of 3.0$\times$ the point-spread function (PSF)'s full-width-half-maximum (FWHM). 
Because of the non-sidereal tracking of our target, the elongated stars in the \name\ images cannot be used to compute a point-source aperture correction. 
We instead derived a mean stellar PSF profile from the calibrator images for each dataset, and used this profile to calculate both the FWHM radius and an aperture correction, for photometry on the non-sidereal \name\ images.
For the optical GMOS photometry, the four bracketing sidereally-tracked images were calibrated to the SDSS system. 
Image zeropoints and a linear color term were fit using the instrumental SDSS magnitudes of stars in the calibration images, to scale the SDSS system to the Gemini filter system.
The zeropoints of the science images were set from those of the calibrator temporally closest to each science image, introducing 0.02 magnitude of uncertainty in the calibration of those frames to encompass the size of the variation in the zeropoint.  
An additional 0.02 magnitude uncertainty is due to the indirect measure of the aperture correction \citep{Fraser:2016a}.
For the J-band NIRI photometry, we apply a mean aperture correction of 0.03 mag on the \name\ data. 
While this does not account for seeing variation along the longer J-band sequence, the highly stable sky conditions during our observations and the use of a large aperture mitigate the need for a variable aperture correction. 
The magnitudes of \name\ in Table~\ref{tab:observations} from the Gemini observations are thus in the Gemini filter system.

Photometry on the WHT imaging (Table~\ref{tab:observations}) used a slightly different analysis, as we did not have sidereally tracked calibrator frames.
The photometry of trailed stars in all images were measured using TRIPPy pill-shaped apertures, with length equal to the known rate of motion during the image.
Stellar centroids were found with Source Extraction and Photometry in Python \citep{Barbary:2017}, using a custom linear kernel of the trail length and angle, which was convolved with a gaussian to simulate the appropriate stellar shape. 
A filter-dependent FWHM of 0.7--1.3\arcsec\ were measured directly from \name. 
It was not visible in the $z$ stack or in the fourth $r$ stack, and thus we do not report $z$ photometry.
Apertures 7 pixels in radius (1.8\arcsec) were used for both the round aperture used to measure the flux from \name, and the pill apertures used on the stars.
As this implicitly assumes identical aperture corrections for both aperture shapes, we adopt 0.02 magnitudes as a conservative estimate of measurement uncertainty, induced by the use of a fixed aperture.
Stellar calibration magnitudes were extracted from the Pan-STARRS1 catalog \citep{Chambers.2016}, and converted to SDSS using the \citet{Tonry:2012} transformations.
As insufficient stars were available to measure a color term, we required the SDSS stars to have similar colors to \name: $0.1<(g-r)<0.7$. 
This induced a 0.02 magnitude uncertainty in calibration.
The magnitudes of \name\ in Table~\ref{tab:observations} from the WHT observations are thus in the SDSS system.

\begin{deluxetable}{rlclc}
\tabletypesize{\footnotesize}
\tablecaption{Photometry of \name\ with the 8.1-m Frederick C. Gillett Gemini North Telescope and the 4.2 m William Herschel Telescope \label{tab:observations}}
\tablehead{\colhead{MJD}\vspace{-0.2cm} & \colhead{Filter}   & \colhead{Effective} & \colhead{$m_{filter}$} & \colhead{Note} \\ \colhead{} & \colhead{} & \colhead{Exposure (s)} & \colhead{} & \colhead{} }
\tablecolumns{5}
\startdata
\cutinhead{GMOS-N and NIRI (Gemini North)}
58055.25860 & r$\_$G0303 & 300 & $22.74 \pm 0.03$ &  \\
58055.26323 & r$\_$G0303 & 300 & $22.66 \pm  0.03$ & \\
58055.26902 & g$\_$G0301 & 300 & $23.11 \pm  0.07$ & \\
58055.27337 & g$\_$G0301 & 300 & $23.02 \pm  0.06$ & \\
58055.28856 & J & 840 & $21.19 \pm 0.07$ & 7-image stack \\
58055.29954	& J	& 840 & $21.29 \pm 0.08$ & 7-image stack \\
58055.30914 & J & 720 & $20.99 \pm 0.07$ & 6-image stack \\
58055.31881 & g$\_$G0301 & 300 & $22.40 \pm 0.03$ & \\
58055.32281 & r$\_$G0303 & 300 & $22.08 \pm 0.02$ & \\
\cutinhead{ACAM (WHT)}
58055.82845 & r$\_$ING702 & 600 & $22.47 \pm 0.09$  & 6-image stack \\
58055.83740 & g$\_$ING701 & 600 & $23.28 \pm 0.30$  & 6-image stack \\
58055.84646 & r$\_$ING702 & 700 & $22.83 \pm 0.11$  & 7-image stack \\
58055.86336 & i$\_$ING703 & 1000 & $22.81 \pm 0.08$  & 10-image stack \\
58055.87835 & r$\_$ING702 & 600 & $23.40 \pm 0.18$  & 6-image stack \\
\enddata
\tablecomments{GMOS-N (Gemini) photometry is color term corrected, in the Gemini-Hamamatsu system. ACAM (WHT) photometry is in the SDSS system. For both, shot noise, calibration, and aperture correction uncertainties are incorporated in quadrature.}
\end{deluxetable}

\subsection{Color Computation}
\label{sec:colours}

For the $g-r$ and $r-J$ colors of \name, a best-fit line and an average $g-r$ color term were fit to the higher-precision Gemini optical data, in the Gemini system (Fig.~\ref{fig:movingJband}). 
A mean $r-J$ color was found by estimating the $r$ value from the fitted line at each J stack epoch. 
Uncertainties in the fit, and in the individual J measurements were folded together.
The $g-r$ and $r-J$ colors thus determined were then converted to the SDSS system using the color terms determined from the initial calibration, carrying uncertainties in the color term appropriately.

The colors from the WHT measurements are consistent, but substantially more uncertain, and we subsequently consider only those from Gemini. 
We note the $r-i$ color corresponds to a spectral slope of $22 \pm 15$\%, consistent with the earlier reports.
The measured SDSS colors of \name\ are in Table~\ref{tab:colors}.

\begin{deluxetable}{ccc}
\tablecaption{Optical-NIR SDSS Colors of \name\ \label{tab:colors}}
\tablehead{ \colhead{Filters}   & \colhead{Measured Color} & \colhead{Observations}}
 \startdata
 $g-r$ & $0.47 \pm 0.04$ & Gemini \\
 $g-r$ & $0.63 \pm 0.31$ & WHT \\
 $r-i$ & $0.36 \pm 0.16$ & WHT \\
 $r-J$ & $1.20 \pm 0.11$ & Gemini \\
\enddata
\tablecomments{Gemini $g-r$ and $r-J$ colors are near-simultaneous; WHT $g-r$ and $r-i$ are from 13 hours later.}
\end{deluxetable}

\section{Variability and Shape of 1I/`Oumuamua}

\begin{figure*}
\plotone{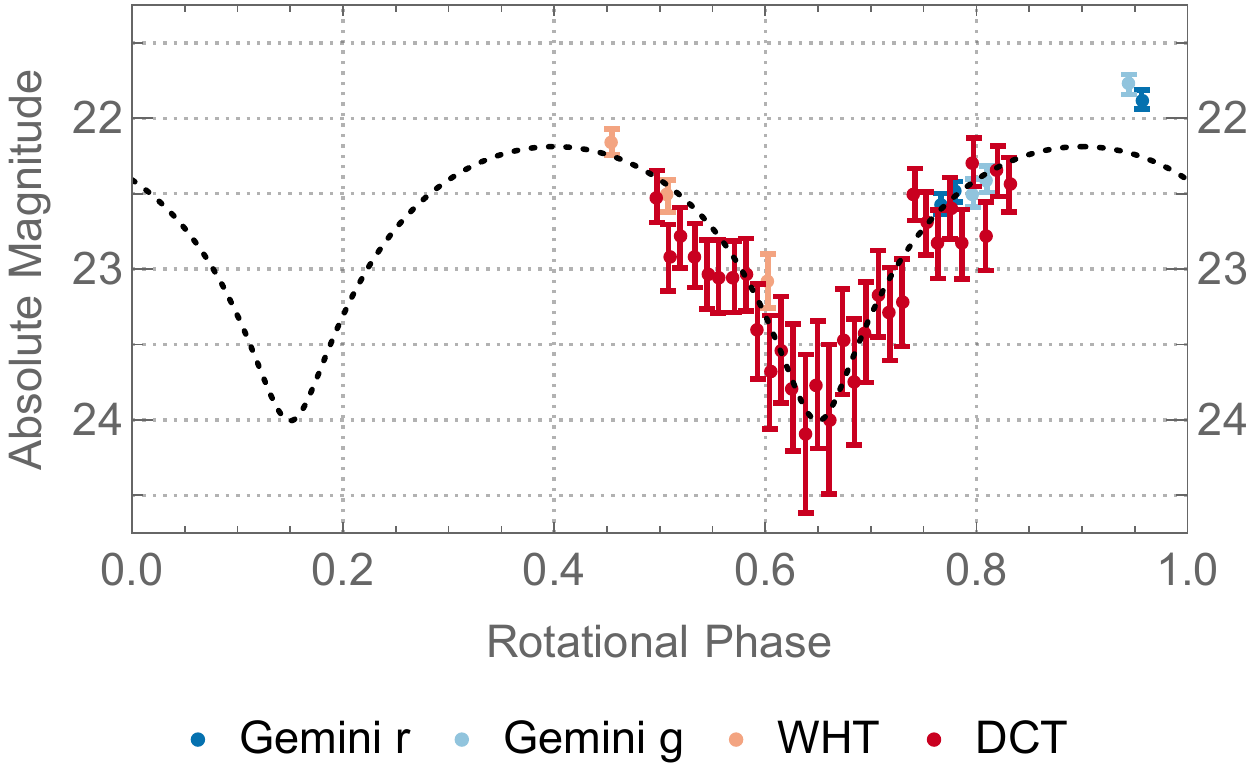}
\caption{Lightcurve of \name\ during 2017 October 29--30. Absolute magnitude assumes a linear phase function with slope $0.03$ mag/deg. Rotational phase starts at MJD 58055.0 and assumes a spin period $P=8.1$ hr. Overplotted is a model based on a prolate ellipsoid with axes ratio $a/b=5.3$, which has $\Delta m=1.8$ mag. A similar curve would be produced by a contact binary with equal sized, prolate components each with $a/b\sim2.7$, elongated along a line connecting their centers.}
\label{fig:lightcurve}
\end{figure*}

There are significant and correlated brightness increases in optical and NIR for \name\ during our Gemini observations, with the variability in J-band tracking that in the r$^\prime$ and g$^\prime$ bands (Fig.~\ref{fig:movingJband}). 
Over the 0.48 hours of J-band imaging, \name\ brightens systematically by $0.183 \pm 0.065$ magnitudes, and likewise, brightens by $0.66 \pm 0.03$ and $0.71 \pm 0.05$ magnitudes between the first and last r' and g' observations, which respectively span 1.54 and 1.19 hours. 
These are significant variations given the short timeframe (Table~\ref{tab:observations}).
As they correlate across filters, the brightening is most likely due to \name's shape rather than to variability in albedo or surface spectral reflectance.
This is supported by the general consistency of the WHT colours, observed at a different part of \name's lightcurve (Fig.~\ref{fig:lightcurve}, discussed below).
Our image stacks all had point-like PSFs, and we thus do not assess upper limits for the presence of coma. 
However, the overall periodicity of \name's brightness (Fig.~\ref{fig:lightcurve}) also implies our observed increase in brightness in the Gemini data is not from dust emission.

We assess the lightcurve of \name\ by combining our optical photometry from 2017 October 29 (Table~\ref{tab:observations}) with the optical photometry of \citet{Knight:2017arXiv} with the 4.3 m Discovery Channel Telescope (DCT) on 2017 October 30. 
The combined, geometrically corrected photometry is in Fig.~\ref{fig:lightcurve}. 
We analyze the photometry for periodicity using both the Lomb-Scargle technique \citep{Lomb:1976} and a modified phase-dispersion minimisation (PDM) fitting technique \citep{Stellingwerf:1978}.
Instead of the typical PDM model which bins the data and looks for the place where the points in the bins are not as dispersed as other periods, our modified PDM goes through every possible period and folds the data, then fits a second-order Fourier series to each folded lightcurve. The quality of fit is calculated from the residuals, after which the best fit is chosen (M. W. Buie, pers. comm.).
We obtain a consistent double-peaked period of $8.10 \pm 0.42$ hours, with a peak-to-peak amplitude from the fitted model of $\Delta m = 1.8$ magnitudes.
We note that the last pair of Gemini observations imply an excursion from our lightcurve model; the data are entirely reliable, given the consistent stability of the observing conditions and the calibrations, and from careful inspection of the images.
Our results using only our Gemini and WHT photometry with that from the DCT are independently in agreement with those of \citet{Bolin:2017}, which used photometry from the 3.5 m Apache Point Observatory and the DCT photometry.

Its lightcurve implies that \name\ is either a very elongated object, or a contact binary system of two equal sized, prolate components aligned for maximum elongation \citep{Sheppard-Jewitt:2004,Leone:1984}. 
From simulations for resolved Centaur binaries in our Solar System \citep{Noll:2006}, a contact binary would stay intact through perihelion.
The light curve yields consistent results in either case.
We consider \name's elongation and density assuming it is a prolate ellipsoid with semi-axes $a>b=c$.
The observed $\Delta m = 1.8$ mags would require an axis ratio of $a/b = 5.3$, or larger if \name\ was not observed equator-on \citep{Lacerda:2003}.
Such an ellipsoid spinning in $P=8.1$ hours with $a/b = 5.3$ (Fig.~\ref{fig:lightcurve}) would require a density at least $\rho = (a/b)^2 (3\pi)/(G P^2)=5.9$ g.cm$^{-3}$ to prevent it from shedding regolith, consistent with the observed absence of coma.
If it is instead a contact binary of two prolate components, each with axes ratio $0.5(a/b)$ (to produce the same $\Delta m$) a similar density of 5.9 g.cm$^{-3}$ is required to hold the components in mutual orbit.
As these densities are unreasonably higher than those of likely compositions of silicate or icy materials, it requires that \name\ has internal strength. 

Small TNOs with $H < 9$ and Centaurs with $H < 11$ typically have 7--9 hour rotation periods with peak-to-peak variations of $\sim$0.3 magnitudes \citep{Duffard:2009}, though light curve amplitude changes of a magnitude or more have been measured for TNOs in this size range \citep{Benecchi:2013}. Small asteroids with \name's degree of elongation are rare but not unknown; examples include the $\sim 200-300$~m diameter Near-Earth Asteroids 2001 FE90 and 2007 MK13, both with lightcurve amplitudes $\geq 2.1$ magnitudes \citep{Warner:2009}.

\section{1I/`Oumuamua in Context with the Colors of the Minor Planets of the Solar System}
\label{sec:discussion}

\begin{figure*}
\plotone{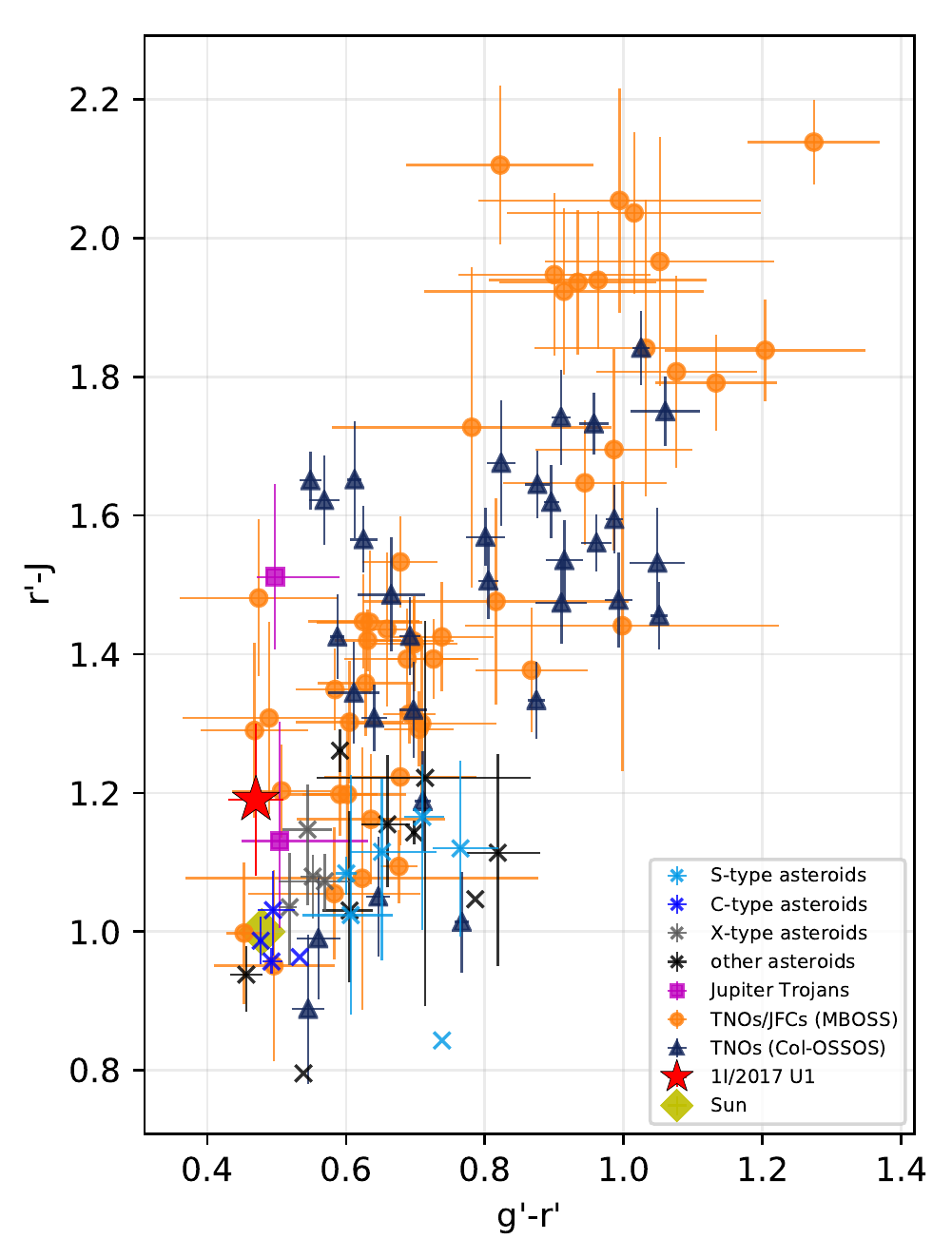}
\caption{The grJ colors of \name\ in context with the known Solar System. The  mean color and range of asteroids and the two color classes of Jupiter Trojans are given from 157 asteroids in 23 of the 26 \citet{Bus:2002taxonomy} spectral types as recorded in SMASS, and 38 Jupiter Trojans \citep{Emery:2011,Marsset:2014}. The much more sparsely sampled distant populations are shown with individual objects, with their measurement uncertainties from MBOSS and Col-OSSOS. The observations of \name\ and of Col-OSSOS were performed in the same way.} 
\label{fig:grJ}
\end{figure*}

For comparison with \name, we collated colors of minor planet populations in our Solar System from four datasets of optical and near-IR measurements: SMASS for asteroids, \citet{Emery:2011,Marsset:2014} for Jupiter Trojans, MBOSS for a variety of distant populations, and Col-OSSOS for trans-Neptunian objects (TNOs).
The Small Main-Belt Asteroid Spectroscopic Survey (SMASS)\footnote{\url{http://smass.mit.edu/smass.html}} \citep{Xu:1995,Burbine:2002,Bus:2002data} measured spectra for a range of asteroid dynamical groups, from near-Earth objects through the main belt to Mars-crossing asteroids. 
SMASS provided 157 objects with both optical and NIR spectra, representing 23 of the 26 spectral types in the \citet{Bus:2002taxonomy} taxonomy (the Cg, D, and Q types were not available in NIR). 
For P and D types, we used the 38 Jupiter Trojans with NIR spectra from \citet{Emery:2011} for which \citet{Marsset:2014} collated optical spectra.
We converted each spectra to grJ colors by convolving the filter bandpasses.
The mean and range for each spectral type are shown in Fig.~\ref{fig:grJ}.
The Minor Bodies in the Outer Solar System (MBOSS) \citep{Hainaut:2012} database\footnote{\url{http://www.eso.org/~ohainaut/MBOSS/}} indexes the reported colors for objects in outer Solar System dynamical populations, including Jupiter Trojans, short- and long-period comets, Centaurs and TNOs. 
47 objects in the MBOSS tabulation have measurements in comparable filters and of sufficiently high-SNR to provide comparison to our grJ measurements of \name.
The MBOSS colors were converted to $g-r$ and $r-J$ assuming a linear spectrum through the $V, R, g, r$ range.
We retained objects if the uncertainties on their color measurements had $d(V-R)<0.4$ and $d(V-J)<0.4$.
Col-OSSOS provides high-precision grJ colors for $m_r < 23.6$ TNOs from the Outer Solar System Origins Survey \citep{Bannister:2016a}, acquired in the same manner as our observations of \name\ with Gemini (\S~\ref{sec:observations}).
We use the 21 Col-OSSOS TNOs with grz photometry discussed in \citet{Pike:2017gf}; the grJ measurements are forthcoming in \citet{Schwamb:2017inprep}.

The grJ colors of \name\ are at the neutral end of the Solar System populations (Fig.~\ref{fig:grJ}). 
About 15\% of the trans-Neptunian objects have colors consistent with \name, all in dynamically excited populations. 
\name's color is also consistent with that of the less red Jupiter Trojans, which are P type \citep{Emery:2011}, and with \citet{Bus:2002taxonomy, DeMeo:2009} X type in the asteroids, which encompasses the \citet{Tholen:1984} E, M and P classifications.
As its albedo is unknown, we do not describe \name\ as consistent with \citet{Tholen:1984} P type. 

Notably, \name\ does not share the distinctly redder colors of the cold classical TNOs \citep{Tegler:2003,Pike:2017gf}, which may be on primordial orbits.
Nor is its color among the red or ``ultra-red" colors of the larger TNOs on orbits that cross or are well exterior to the heliopause \citep{Sheppard:2010wy,Trujillo:2014ih,Bannister:2017OSSOSV}.
The cause of ultra-red coloration of these TNOs is unknown, but  has been attributed to long-term cosmic ray alteration of organic-rich surfaces \citep{Jewitt:2002}, such as would be expected during the long duration of interstellar travel.

While this work was under review, several other well-constrained color and spectral measurements were reported.
Our optical color is compatible with that observed in $BVR$ by \citet{Jewitt:2017}, but $3\sigma$ discrepant from the mean $g-r = 0.84 \pm 0.05$ over multiple nights of \citet{Meech:2017}.
Our measurements confirm consistent color over  roughly a quarter of \name's surface, so the discrepancy likely indicates a change in surface color elsewhere on the body.
This is supported by considering spectral slope through the near-IR; our $r-J$ corresponding to 3.6\%/100nm is more neutral than the $7.7 \pm 1.3$\%/100nm observed by \citet{Fitzsimmons:2017} over $0.63\mu$m to $1.25\mu$m.
Note that our 1.15--1.33$\mu$m J-band data is at longer wavelengths than the \citet{Meech:2017} $Y$-band (0.97--1.07$\mu$m), so we make no direct comparison there.
More extensive modelling of surface color patchiness and non-principal axis rotation (tumbling) of \name\ is considered by \citet{Fraser:2017}.

\name's largely neutral color opens up a number of possibilities.
It could imply that the correlation of ultra-redness with heliocentric distance has an alternative cause.
It could suggest that \name\ formed with an organics-poor surface, within its star's water ice line. 
\name's color being within the observed range for minor planets in the Solar System could support that \name\ originated from a star from the Sun's birth cluster, which should have a similar chemistry.
A possible additional complication could be resurfacing due to surface activity, which would affect surface color. 
This seems unlikely as no surface activity was detected during \name's perihelion passage, but \name\ could have had past activity in its origin system or in another close encounter.
We emphasize that our observations only probe the top few microns of \name's surface.

Our Gemini and WHT observations provide high-precision, lightcurve-independent optical and NIR color measurements for \name.
With its period of $8.1 \pm 0.4$ hours, highly elongated $\geq 5.3:1$ ellipsoidal or prolate-binary shape, and neutral $grJ$ color, \name\ is within the known parameters of minor planets from the Solar System, but lies at the extreme ends of the physical ranges.

\acknowledgments

The authors acknowledge the sacred nature of Maunakea and appreciate the opportunity to observe from the mountain.  
This work is based on observations from Director's Discretionary Time (DDT) Program GN-2017B-DD-8, obtained at the Gemini Observatory, which is operated by the Association of Universities for Research in Astronomy, Inc., under a cooperative agreement with the NSF on behalf of the Gemini partnership: the National Science Foundation (United States), the National Research Council (Canada), CONICYT (Chile), Ministerio de Ciencia, Tecnolog\'{i}a e Innovaci\'{o}n Productiva (Argentina), and Minist\'{e}rio da Ci\^{e}ncia, Tecnologia e Inova\c{c}\~{a}o (Brazil). The authors thank the queue coordinators,  NIRI and GMOS instrument teams, science operations specialists, and other observatory staff at Gemini North for their support of our DDT program and their support of the Col-OSSOS program. We also thank Andy Stephens for his assistance during the Gemini observations.

The WHT is operated on the island of La Palma by the Isaac Newton Group of Telescopes in the Spanish Observatorio del Roque de los Muchachos of the Instituto de Astrof\'{i}sica de Canarias. The ACAM data were obtained as part of programme SW2017b11. We thank Ian Skillen for advising on and performing the WHT observations.

The authors greatly thank Atsuko Nitta for her support of the Gemini observing program and for being a sounding board for program ideas and observing strategies. We especially acknowledge and thank the online planetary community on Twitter for productive discourses and sharing of preliminary results related to \name\ that helped spur these observations. 

M.T.B. appreciates support from UK STFC grant ST/L000709/1. M.E.S., A.B.S. and S.S. were supported by Gemini Observatory. 

This research has made use of NASA's Astrophysics Data System Bibliographic Services, the JPL HORIZONS web interface (\url{https://ssd.jpl.nasa.gov/horizons.cgi}), and data and services provided by the International Astronomical Union's Minor Planet Center. 

\vspace{5mm}
\facilities{Gemini:Gillett (GMOSN, NIRI), ING:Herschel (ACAM)}

\software{astropy \citep{TheAstropyCollaboration:2013cd},
		  TRIPPy \citep{Fraser:2016a},
          SExtractor \citep{Bertin:1996},
          SEP \citep{Barbary:2017},
          AstroConda (\url{http://astroconda.readthedocs.io/},
          L.A. Cosmic \citep{vanDokkum:2001})
          }

\bibliography{references}

\end{document}